\documentclass[twocolumn]{aastex631}

\usepackage{CJKutf8}
\usepackage{amsmath}
\usepackage{soul}

\newcommand{\TQ}{$t_{Q}$}

\newcommand{\oiiitext}{[O~{\sc iii}]}

\newcommand{\lya}{Ly$\alpha$}

\newcommand{\cii}{[C~{\sc ii}] 158 \mum}
\newcommand{\ciitext}{[C~{\sc ii}]}

\newcommand{\mum}{\ifmmode{\rm \mu m}\else{$\mu$m}\fi}
\newcommand{\vdisp}{$\vdisp$}

\newcommand{\vwu}{{v$_{50}$}}

\newcommand{\oiii}{{[O$\,${\scriptsize III}] $\lambda$5007}}
\newcommand{\oiiiab}{{[O$\,${\scriptsize III}] $\lambda$$\lambda$4959,5007}}

\newcommand{\civ}{\hbox{C$\,${\scriptsize IV} $\lambda$1549}}

\newcommand{\hb}{\hbox{H$\beta$}}
\newcommand{\hg}{\hbox{H$\gamma$}}

\newcommand{\mgii}{\hbox{Mg$\,${\scriptsize II}}}
\newcommand{\feii}{\hbox{Fe$\,${\scriptsize II}}}

\newcommand{\kms}{km s$^{-1}$} 
\newcommand{\msun}{M$_{\odot}$} 
\newcommand{\msunyr}{{M$_{\sun}$ yr$^{-1}$}}

\newcommand{\ergs}{erg\,s$^{-1}$}

\newcommand\qtdfit{\texttt{q3dfit}}

\shorttitle{J1007$+$2115}
\shortauthors{Liu al.}
\graphicspath{{./}{figures/}}

\begin{document}



\title{Fast Outflow in the Host Galaxy of the Luminous $z=7.5$ Quasar J1007$+$2115}

\author[0000-0003-3762-7344]{Weizhe Liu \begin{CJK}{UTF8}{gbsn}(刘伟哲)\end{CJK}}
\affiliation{Steward Observatory, University of Arizona, 933 N. Cherry Ave., Tucson, AZ 85721, USA}

\correspondingauthor{Weizhe Liu}
\email{wzliu@arizona.edu}

\author[0000-0003-3310-0131]{Xiaohui Fan}
\affiliation{Steward Observatory, University of Arizona, 933 N. Cherry Ave., Tucson, AZ 85721, USA}

\author[0000-0001-5287-4242]{Jinyi Yang}
\affiliation{Steward Observatory, University of Arizona, 933 N. Cherry Ave., Tucson, AZ 85721, USA}

\author[0000-0002-2931-7824]{Eduardo Ba\~nados}
\affil{Max Planck Institut f\"ur Astronomie, K\"onigstuhl 17, D-69117, Heidelberg, Germany}

\author[0000-0002-7633-431X]{Feige Wang}
\affiliation{Steward Observatory, University of Arizona, 933 N. Cherry Ave., Tucson, AZ 85721, USA}

\author[0000-0003-0643-7935]{Julien Wolf}
\affil{Max Planck Institut f\"ur Astronomie, K\"onigstuhl 17, D-69117, Heidelberg, Germany}

\author[0000-0002-3026-0562]{Aaron J. Barth}
\affil{Department of Physics and Astronomy, University of California, Irvine, CA 92697, USA}

\author[0000-0002-6748-2900]{Tiago Costa}
\affiliation{School of Mathematics, Statistics and Physics, Newcastle University, Newcastle upon Tyne, NE1 7RU, UK}

\author[0000-0002-2662-8803]{Roberto Decarli}
\affil{INAF - Osservatorio di Astrofisica e Scienza dello Spazio di Bologna, via Gobetti 93/3, I-40129, Bologna, Italy}

\author[0000-0003-2895-6218]{Anna-Christina Eilers}
\affiliation{Department of Physics, Massachusetts Institute of Technology, Cambridge, MA 02139, USA}
\affiliation{MIT Kavli Institute for Astrophysics and Space Research, Massachusetts Institute of Technology, Cambridge, MA 02139, USA}

\author[0000-0002-8858-6784]{Federica Loiacono}
\affil{INAF - Osservatorio di Astrofisica e Scienza dello Spazio di Bologna, via Gobetti 93/3, I-40129, Bologna, Italy}

\author[0000-0003-1659-7035]{Yue Shen}
\affiliation{Department of Astronomy, University of Illinois at Urbana-Champaign, Urbana, IL 61801, USA}
\affiliation{National Center for Supercomputing Applications, University of Illinois at Urbana-Champaign, Urbana, IL 61801, USA}

\author[0000-0002-6822-2254]{Emanuele Paolo Farina}
\affiliation{International Gemini Observatory/NSF NOIRLab, 670 N A’ohoku Place, Hilo, Hawai'i 96720, USA}

\author[0000-0002-5768-738X]{Xiangyu Jin}
\affiliation{Steward Observatory, University of Arizona, 933 N. Cherry Ave., Tucson, AZ 85721, USA}

\author[0000-0003-1470-5901]{Hyunsung D. Jun}
\affiliation{Department of Physics, Northwestern College, 101 7th St SW, Orange City, IA 51041, USA}

\author[0000-0001-6251-649X]{Mingyu Li}
\affiliation{Department of Astronomy, Tsinghua University, Beijing 100084, People's Republic of China}

\author[0000-0001-6106-7821]{Alessandro Lupi}
\affil{Dipartimento di Scienza e Alta Tecnologia, Universit\`a degli Studi dell'Insubria, via Valleggio 11, I-22100, Como, Italy}
\affil{INFN, Sezione di Milano-Bicocca, Piazza della Scienza 3, I-20126 Milano, Italy}
\affil{Dipartimento di Fisica ``G. Occhialini'', Universit\`a degli Studi di Milano-Bicocca, Piazza della Scienza 3, I-20126 Milano, Italy}

\author[0000-0001-6434-7845]{Madeline A. Marshall}
\affil{National Research Council of Canada, Herzberg Astronomy \& Astrophysics Research Centre, 5071 west Saanich Road, Victoria, BC V9E 2E7, Canada}
\affil{ARC Centre of Excellence for All Sky Astrophysics in 3 Dimensions (ASTRO 3D), Australia}

\author[0000-0003-0230-6436]{Zhiwei Pan}
\affiliation{Kavli Institute for Astronomy and Astrophysics, Peking University, Beijing 100871, China}
\affiliation{Department of Astronomy, School of Physics, Peking University, Beijing 100871, China}

\author[0000-0003-4924-5941]{Maria Pudoka}
\affiliation{Steward Observatory, University of Arizona, 933 N. Cherry Ave., Tucson, AZ 85721, USA}

\author[0000-0001-5105-2837]{Ming-Yang Zhuang \begin{CJK}{UTF8}{gbsn}(庄明阳)\end{CJK}}
\affiliation{Department of Astronomy, University of Illinois at Urbana-Champaign, Urbana, IL 61801, USA}

\author[0000-0002-6184-9097]{Jaclyn~B.~Champagne}
\affiliation{Steward Observatory, University of Arizona, 933 N. Cherry Ave., Tucson, AZ 85721, USA}

\author{Huan Li}
\affiliation{School of Aerospace Science and Technology, Xidian Univerisity, Xian, Shaanxi, 710126 China}

\author[0000-0002-4622-6617]{Fengwu Sun}
\affiliation{Steward Observatory, University of Arizona, 933 N. Cherry Ave., Tucson, AZ 85721, USA}

\author[0000-0003-0747-1780]{Wei Leong Tee}
\affiliation{Steward Observatory, University of Arizona, 933 N. Cherry Ave., Tucson, AZ 85721, USA}

\author[0000-0002-0710-3729]{Andrey Vayner}
\affiliation{IPAC, California Institute of Technology, 1200 E. California Boulevard, Pasadena, CA 91125, USA}

\author[0000-0002-4321-3538]{Haowen Zhang}
\affiliation{Steward Observatory, University of Arizona, 933 N. Cherry Ave., Tucson, AZ 85721, USA}




\begin{abstract}
James Webb Space Telescope opens a new window to directly probe luminous quasars powered by billion solar mass black holes in the epoch of reionization and their co-evolution with massive galaxies with unprecedented details. In this paper, we report the first results from the deep NIRSpec integral field spectroscopy study of a quasar at $z = 7.5$. We obtain a bolometric luminosity of $\sim$$1.8\times10^{47}$ \ergs\ and a black hole mass of {$\sim$0.7--2.5$\times10^{9}$} \msun\ based on \hb\ emission line from the quasar spectrum. We discover $\sim$2 kpc scale, highly blueshifted ($\sim$$-$870 \kms) and broad ($\sim$1400 \kms) \oiiitext\ line emission after the quasar PSF has been subtracted. Such line emission most likely originates from a fast, quasar-driven outflow, the earliest one on galactic-scale known so far. The dynamical properties of this outflow fall within the typical ranges of quasar-driven outflows at lower redshift, and the outflow may be fast enough to reach the circumgalactic medium. Combining both the extended and nuclear outflow together, the mass outflow rate, {$\sim$300 \msunyr, is $\sim$60\%--380\%} of the star formation rate of the quasar host galaxy, suggesting that the outflow may expel a significant amount of gas from the inner region of the galaxy. The kinetic energy outflow rate, {$\sim$3.6$\times10^{44}$ erg s$^{-1}$}, is {$\sim$0.2\%} of the quasar bolometric luminosity, which is comparable to the minimum value required for negative feedback based on simulation predictions. The dynamical timescale of the extended outflow is $\sim$1.7 Myr, consistent with the typical quasar lifetime in this era. 


\end{abstract}



\section{Introduction} 
\label{sec:intro}

Luminous quasars powered by $\sim$$10^9$ $M_\odot$ black holes already exist in the Epoch of Reionization (EoR), raising the open question of how such massive systems form so rapidly. \citep[e.g.,][and references therein]{Volonteri2012,Wu2015,Banados2018,Matsuoka2019b,Onoue2019,Shen2019,Yang2020b,Schindler2020,Yang2023b,Fan2023,Mazzucchelli2023,Bigwood2024}. Sensitive sub-mm observations have revealed rapid star formation and large amount of cool gas and dust in the host galaxies of these quasars, and depicted the diverse gas kinematics and dynamics in them \citep[e.g.,][and references therein]{Willott2015,venemans_kiloparsec-scale_2020,Neeleman_kinematics_2021,Izumi2021}. While the detection of stellar components from their host galaxies remained extremely challenging in the Hubble Space Telescope (HST) era \citep[e.g.,][]{Mechtley2012,Marshall2020}, we are finally able to unveil them in several such quasar host galaxies with the advent of JWST. The emerging results already paint a complex picture: while some suggest overmassive black holes with respect to their host galaxies when compared to the local scaling relations \citep[e.g.][Yang et al. in prep.]{yue_eiger_2023,Stone2023}, measurements in lower luminosity quasars suggests that they are consistent with the local scaling relation \citep{Ding2023}.

In addition to the stellar components, gaseous environments of these high-$z$ quasars are also expected to be complex based on both observations and simulations \citep[e.g.,][]{Li2007,DiMatteo2012,Ni2018,Lupi2019,Lupi2022,Costa2022,farina_requiem_2019}: The quasars and their host galaxies are fueled by cold gas streams and gas-rich mergers, while powerful outflows driven by the quasars and starburst activities are transporting mass, momentum, and energy outwards. Careful observations are needed to closely scrutinize predictions from these simulations. 
In particular, while detections of nuclear quasar winds and galactic-scale quasar-driven outflows are rapidly accumulating, such observations remain challenging and sometimes lead to contradictory results:
On parsec scale, quasar winds traced by blueshifted and/or broad rest-frame UV emission and absorption lines have been reported repeatedly \citep[e.g.,][]{Meyer2019,Schindler2020,Yang2021,Bischetti2022}. On the galactic scale, while a powerful quasar-driven outflow at $z\sim6.4$ has been reported in SDSS J114816.64$+$525150.3 \citep{maio12, Cicone2015}, \citet{Meyer2022}, instead, find no evidence of outflow in the same source. Likewise, both \citet{Stanley2019} and \citet{Bischetti2019} report widespread quasar-driven outflows in the early universe by stacking the \cii\ emission in a sample of 20 quasars at $z\sim6$ and a sample of 45 quasars at $4.5<z<7$.1, respectively. However, a study of 27 quasars at $z\gtrsim6$ from \citet{Novak2020}, adopting a different stacking technique, argues for no evidence of fast \cii\ outflows in their sample. More recently, initial results from the JWST\ program ``A SPectroscopic survey of biased halos In the Reionization Era'' (ASPIRE) \citep{wang_spectroscopic_2023} revealed fast outflows in several quasars at $z>6.5$ through broad and blueshifted \oiii\ emission line, adding new tantalizing evidence for quasar feedback in the early universe \citep{Yang2023b}. Modern simulations suggest that quasar feedback via powerful outflows is already at work at $z\gtrsim6$ \citep[e.g.,][]{Costa2018,Ni2018,Costa2022}. Some studies \citep[e.g.,][]{Hartley2023,Lovell2023} suggest that such feedback may explain the existence of massive, quenched/quenching galaxies at $z\sim$2-5 \citep[e.g.,][]{Labbe2005, Glazebrook2017, Valentino2020}, and may also be required to reproduce the observed distribution of galaxy masses at $z=0$ \citep[e.g.,][]{Kaviraj2017}.

The NIRSpec integral field unit (IFU) onboard JWST\ provides us with the new opportunity to spatially resolve the gaseous nebulae in and around these high-$z$ quasars through rest-frame optical line emission \citep[e.g.][Decarli et al. under review; Lyu et al. in prep.; Wolf et al. in prep.]{Marshall2023,Loiacono2024}, revealing unprecedented details of the interstellar medium (ISM) and outflows within host galaxies and close companions near quasars.
In this paper, we adopt deep NIRSpec IFU observations of one of the highest redshift quasars known today, J100758.26$+$211529.2 at $z=7.5149$, from the JWST\ cycle 1 Program ``A Comprehensive JWST View of the Most Distant Quasars Deep Into the
Epoch of Reionization'' (ID 1764; PI X. Fan), to study the extended line emission within its host galaxy. This JWST\ program
is dedicated to obtaining a comprehensive view of the three highest redshift quasars known today at $z>7.5$ with NIRCam imaging, NIRSpec fixed-slit and IFU spectroscopy, and MIRI imaging and spectroscopy. 

J100758.26$+$211529.2 (J1007$+$2115 hereafter) was discovered by \citet{Yang2020b} at $z=7.5149$ based on the \cii\ emission line, with a bolometric luminosity of $(2.04\pm{0.13})\times10^{47}$ erg s$^{-1}$ based on 3000 \AA\ continuum luminosity, a black hole mass of $(1.43\pm{0.22})\times10^{9}$ \msun\ based on \mgii\ emission, and a derived Eddington ratio of $1.1\pm{0.2}$ \citep{Yang2021}. The \civ\ emission line is highly blueshifted ($\sim -5000$ \kms) with respect to the systemic velocity based on \cii, which may be interpreted as a fast quasar-driven nuclear wind \citep{Yang2021}. The host galaxy of this quasar has abundant molecular gas ($(2.2\pm{0.2})\times10^{10}$ \msun) and dust \citep[$(1.7\pm{0.6})\times10^{8}$ \msun;][]{Feruglio2023}, and is experiencing rapid star formation, with a star formation rate (SFR) of 80$-$520 \msunyr based on \cii\ luminosity \citep{Yang2020b}.

The rest of the paper is organized as follows: In Section \ref{sec:2}, we describe the JWST\ NIRSpec observations and data reduction. 
In Section \ref{sec:3}, we discuss in detail our methods for subtracting the quasar PSF and measuring the emission from the host galaxy. The obtained emission line properties are then presented in Section \ref{sec:4}. In Section \ref{sec:5}, we further discuss the properties and impact of the outflow traced by the line emission detected. Finally, a brief summary of our findings is presented in Section \ref{sec:6}.
Throughout the paper, we assume a $\Lambda$CDM cosmology with $H_0 =$ 70 km s$^{-1}$ Mpc$^{-1}$, $\Omega_{\rm m}$ = 0.3, and $\Omega_{\rm \Lambda} = 0.7$. An arcsecond corresponds to 5.007 kpc at the redshift of our object.

\section{Observations and Data Reduction} 
\label{sec:2}

\subsection{Observation}
\label{subsec:21}

J1007$+$2115 was observed on Nov 17, 2022 by JWST\ with the NIRSpec/IFU \citep{Bok2022, Jak2022}. These data are available on the Mikulski Archive for Space Telescopes (MAST) at the Space Telescope Science Institute, which can be accessed via \dataset[10.17909/65h6-2671]{http://dx.doi.org/10.17909/65h6-2671}. The IFU observation has a field of view (FOV) of $\sim 3\arcsec \times 3 \arcsec$. We adopted a grating/filter configuration of G395M/F290LP and a NRSIRS2 readout pattern, with a wavelength coverage of $\sim$2.87--5.10~$\mu$m or $\sim$3371--5990\AA\ at the redshift of J1007$+$2115. The grating has a nominal resolving power $\lambda / \Delta\lambda \simeq 1000$, corresponding to a velocity resolution $\sim 300$ \kms, which allows us to spectrally resolve emission lines profiles with typical velocity widths of several hundred \kms. We adopted an 8-point small cycling dither pattern to improve the spatial sampling and enable a good characterization of the point spread function (PSF). We chose the NRSINRS2 readout pattern with 20 groups per integration and 3 integrations per dither. Additionally, one ``leakcal'' exposure at the last dither position was taken to account for light leaking through the closed micro-shutter array (MSA) and from the failed open shutters. The total exposure time was 9.82 hours on source and 1.23 hours for the leakage exposure. 
In addition to the quasar, a nearby PSF star was also observed for a 0.17-hour exposure, with the same instrument set-up, except for a NRSRAPID readout pattern to avoid saturation. While the original request was to observe the PSF star right after the quasar, the first attempt for the PSF star of J1007$+$2115 failed and the PSF star was re-observed a year later on Nov. 14, 2023.

\subsection{Data Reduction}
\label{subsec:22}

The NIRSpec data of J1007$+$2115 
were reduced following the general steps of JWST\ Science Calibration Pipeline (version ``11.12.3'' and context file ``jwst 1019.pmap''), combined with customized software and scripts to replace or improve certain steps in the public pipeline and produce the final data cube properly.

The first stage of the pipeline, \verb|Detector1Pipeline|, performs standard infrared detector reduction steps,  including group scale correction, saturation check, super bias subtraction, reference pixel correction, linearity, persistence correction, ramp Jump detection, fitting ramps of non-destructive group readouts, and gain scale correction.
In this stage, in addition to the pipeline default steps, we also use the \textit{snowblind} software\footnote{\url{https://github.com/mpi-astronomy/snowblind}} to remove noise features caused by ``snowball and shower'' artifacts\footnote{\url{https://jwst-docs.stsci.edu/data-artifacts-and-features/snowballs-and-shower-artifacts}}. We further subtract the 1/$f$ noise in the count rate images, where the correlated vertical noise in each column (i.e. along the spatial axis) is modeled with a 2nd-order polynomial function, after all bright pixels associated with the observed target have been removed through sigma-clipping. 

Next, in the second stage, \verb|Spec2Pipeline|, we apply world coordinate system assignment, flat field correction, and flux calibration. For each individual exposure frame, the 2D spectra are then converted into a 3D data cube using the \verb|cube build| routine, where we adopt an ``EMSM-weighting''\footnote{See Section ``Weighting'' in \url{https://jwst-pipeline.readthedocs.io/en/latest/jwst/cube_build/main.html}} to suppress the spectral oscillations in spectra extracted from individual spaxels caused by the undersampling of the PSF \citep[see][for more detailed illustrations]{Law2023,Perna2023}, at the cost of mild degradation in the spatial and spectral resolution. Additionally, we have skipped the imprint subtraction step as it introduces extra noise to the data. 

Finally, we use customized scripts built upon the Python Package \verb|reproject|\footnote{\href{https://reproject.readthedocs.io/en/stable/}{https:reproject.readthedocs.io/en/stable/}} to generate the final combined data cube: first, a sigma-clipping routine is applied across the eight individual dithers to detect and reject outliers. Subsequently, a common WCS system is determined to which each individual dithered data cube is reprojected, adopting the flux-conserving “reproject\_exact” routine in \verb|reproject|. The final combined data cube is chosen to have a spatial pixel scale of 0\farcs05.

\begin{figure*}[!htb]
\epsscale{1.2}
\plotone{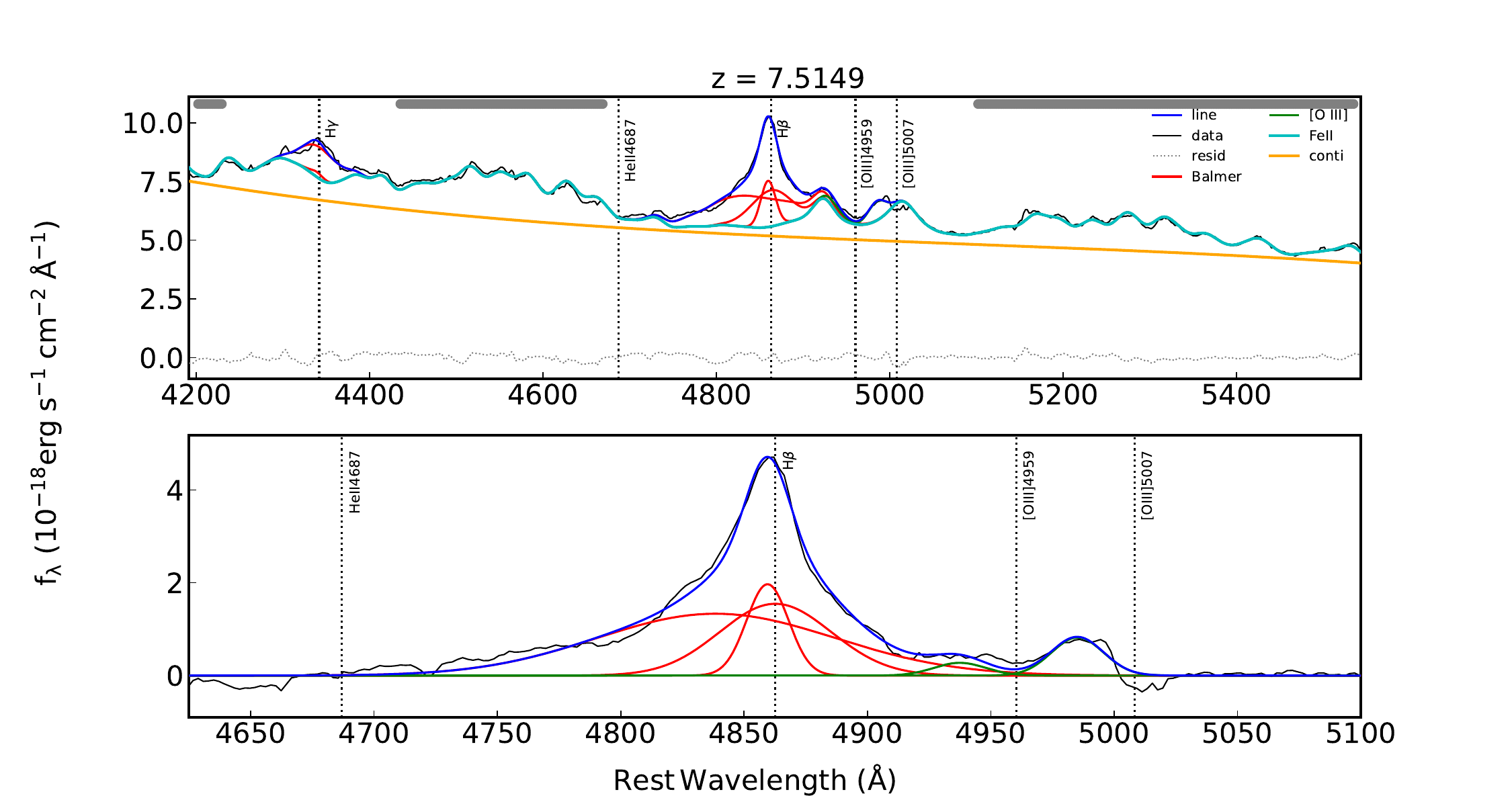}
\caption{Quasar spectrum extracted from the $r=0\farcs25$ aperture centered on the brightest spaxel of the IFU data cube. \textbf{Top:} Data (black) and best-fit quasar continuum (orange), iron emission (cyan) and overall emission line model (blue). The individual Balmer and \oiiitext\ emission line components are shown in red and green, respectively. The locations of major emission lines at the systemic velocity are indicated by the vertical dotted lines. The residual from the best-fit is shown in the dotted, light gray line. The spectral windows adopted for quasar pseudo-continuum fit are marked in thick gray bars. \textbf{Bottom:} Zoom-in view of the emission line-only spectrum for the \hb\ and \oiiitext\ region. The color coding is the same as the top panel.}
\label{fig:fig0}
\end{figure*}

\section{A Quick Look at the quasar spectral properties}\label{sec:3}

The quasar spectrum extracted from an aperture with 0\farcs25 radius centered on the brightest spaxel of the IFU data is shown in Fig.\ \ref{fig:fig0}. This aperture size is large enough to cover $\sim$2.5 times the FWHM of the PSF at the wavelength of the \hb\ emission line while excluding all low S/N spaxels. Broad \hb\ and (weaker) \hg\ emission lines arising from the broad line region (BLR) are detected. Iron emission is prominent, and weak \oiiiab\ doublet is detected based on the spectral fitting described in the next paragraph. The strong iron emission and weak \oiiitext\ emission suggest that the quasar has a high Eddington ratio \citep[e.g.,][]{BorosonGreen1992,ShenHo2014}. The power law continuum from the quasar is also prominent in the spectrum.

We fit the quasar spectrum in Fig.\ \ref{fig:fig0} using the public software, PyQSOFit \citep{pyqsofit,Shen2019}. The quasar pseudo-continuum is fit with a power-law, low-order polynomials, and empirical \feii\ templates from \citet[][]{BorosonGreen1992,Vestergaard2001}, using the continuum windows free of strong quasar emission lines (as indicated by gray horizontal bars in Fig. \ref{fig:fig0}). The emission line-only spectrum is then obtained by subtracting the best-fit pseudo-continuum from the original spectrum. For emission lines, the \hb\ and \hg\ are both fit with three Gaussian components where the kinematics (i.e., velocity and velocity dispersion) of each corresponding Gaussian component in the two lines are tied together. The \oiiitext\ doublet is fit with up to two Gaussian components, but only one Gaussian component is required by the best-fit model based on Bayesian information criterion. The flux ratio of the doublet is fixed at 1:2.98 \citep{Osterbrock2006} in the fit. In addition, we also tried tying the kinematics of one Gaussian component of \hb\ and \hg\ to that of \oiiitext, but the fit resulted in significant residuals blueward of the location of \oiiitext\ emission at systemic velocity and reduced $\chi^2 \gg 1$. Therefore, in our final best-fit model, the \hb\ and \hg\ and the \oiiitext\ doublet are kept as independent components. {The corresponding \hb\ velocity component sharing the same kinematics of \oiiitext\ is not detected.} 


An aperture correction is needed for flux measurements, and we examine the wavelength dependence of the aperture loss by comparing the spectra extracted from various apertures. We find that for apertures with r$\geq$0\farcs25, the change in continuum slope becomes trivial ($\lesssim$1\%). This suggests that the wavelength dependence of the aperture correction is negligible for our chosen aperture size.  
We thus derive the aperture correction for the quasar spectrum above, by comparing the flux within the 0\farcs25 aperture and the total flux (within a radius of 1\farcs5) from the monochromatic image at the peak \hb\ wavelength. The correction obtained is an increase of 16\% in flux, which is then applied to all flux-based measurements.

From our best-fit, we obtain a 5100\AA\ continuum luminosity of $(1.92\pm{0.04}) \times10^{46}$ \ergs, which leads to a bolometric luminosity of $(1.8\pm{0.1}) \times 10^{47}$ \ergs\ based on the 5100\AA\ continuum luminosity following \citet{Richards2006}. The black hole mass is derived adopting the scaling relation
in \citet{Vestergaard2006}:

\begin{equation}
\begin{split}
{\rm log}(M_{BH}) = {\rm log}\left\{[\frac{{\rm FWHM}(H\beta)}{1000\ {\rm km\ s}^{-1}}]^2 [\frac{\lambda L_\lambda(5100 )}{10^{44}\ {\rm erg\ s}^{-1}}]^{0.50}
\right\} \\
+ (6.91 \pm{0.02})
\end{split}
\label{eq:BH}
\end{equation}
{In previous studies, Gaussian components of \hb\ with FWHM larger than 1200 \kms\ are considered part of the emission originating from the BLR and included in the calculation of the FWHM(\hb) above to derive black hole mass \citep[e.g.,][]{ShenLiu2012,Yang2023b}. Here FWHM(\hb) is calculated for the overall profile consisting of all Gaussian components that are considered part of the BLR emission. The narrowest component of \hb\ from our best-fit has a FWHM of 1240$\pm{10}$ \kms, just above the 1200 \kms\ threshold. While this component should be formally considered part of the BLR emission, the moderate spectral resolution (R $\sim$ 1000) leaves open the possibility that an even narrower, non-BLR component could be unresolved and hidden within the component with FWHM $\sim$ 1240 \kms. Therefore, it is debatable whether the 1240 \kms\ component should be included in the calculation of FWHM(\hb). To reflect the uncertainty of FWHM(\hb) caused by this, we calculate the black hole mass adopting two approaches:} (1) By combining all three Gaussian components, we obtain a FWHM(\hb) of $2400\pm{20}$ \kms\ for the entire \hb\ profile and thus a black hole mass of $(7.0\pm{0.4})\times 10^{8}$ \msun\ and an Eddington ratio of 2.2$\pm{0.1}$. (2) Instead, if we exclude the narrowest Gaussian component (i.e., the one with FWHM $\sim$ 1240 \kms), we obtain a FWHM(\hb) of $4730\pm{20}$ \kms\ for the entire \hb\ profile and thus a black hole mass of $(2.5\pm{0.2})\times 10^{9}$ \msun\ and an Eddington ratio of 0.6$\pm{0.1}$. Additionally, the total \hb\ luminosity based on our best-fit is $(2.4\pm{0.1})\times10^{44}$ erg s$^{-1}$.

Comparing with the results from \citep{Yang2021}, the bolometric luminosity derived here based on 5100 \AA\ continuum luminosity is consistent with that based on 3000 \AA\ continuum luminosity, which is $(2.04\pm{0.13})\times10^{47}$ erg s$^{-1}$. {The black hole mass based on \hb\footnote{Note that for quasars with high Eddington ratios like our object, single-epoch mass estimators may significantly overestimate the BH mass based on low-z studies \citep[up to a factor of few; e.g.,][]{Du2019}.}, as derived here, is also broadly consistent with that based on \mgii\ emission, $(1.43\pm{0.22})\times10^{9}$ \msun, given further the typical scatter of $\sim$0.5 dex in the scaling relations for black hole mass measurements like Eq. \ref{eq:BH} above \citep[e.g.,][]{Vestergaard2006}.} 
A more comprehensive discussion on the implication of black hole properties derived above, as well as other quasar properties, will be presented in a future work.

 {Based on our best-fit, the \oiii\ emission line is highly blueshifted ($-1380\pm{20}$ \kms), with a FWHM of 1480$\pm{20}$ \kms\ and luminosity of $(1.77\pm{0.07})\times10^{43}$ erg $s^{-1}$.  Throughout the paper, the systemic velocity is based on \cii\ \citep{Yang2020b}, which is usually a reliable tracer of the kinematically quiescent interstellar medium \citep[e.g.,][]{venemans_kiloparsec-scale_2020}. This is further supported by the narrow \ciitext\ line width of our object \citep[FWHM$\sim$330 \kms;][]{Yang2020b}  }

\begin{figure*}
\plottwo{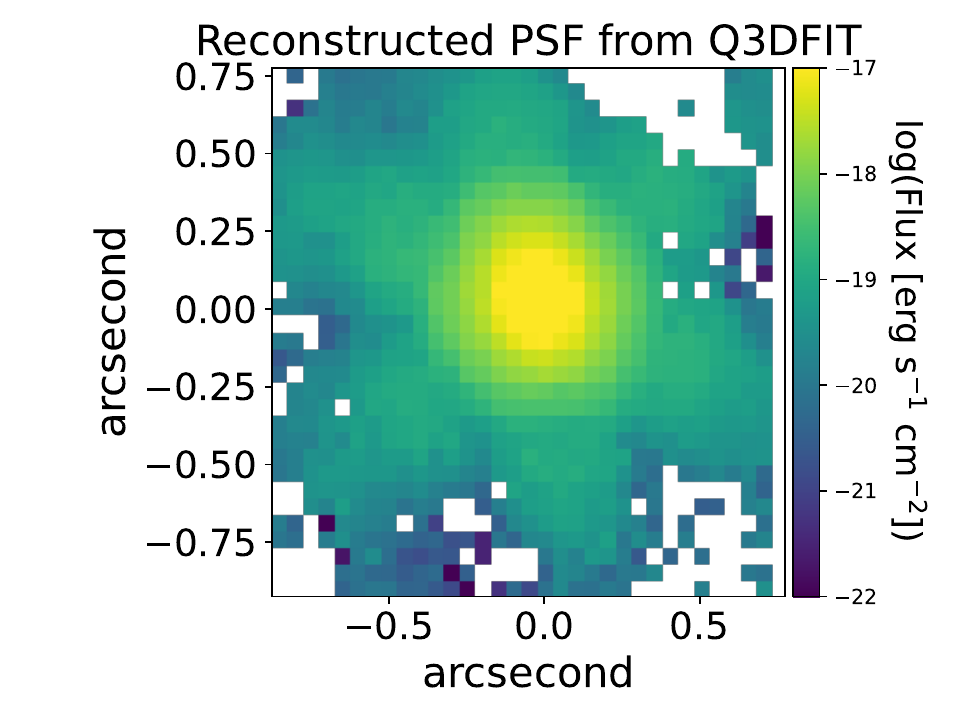}{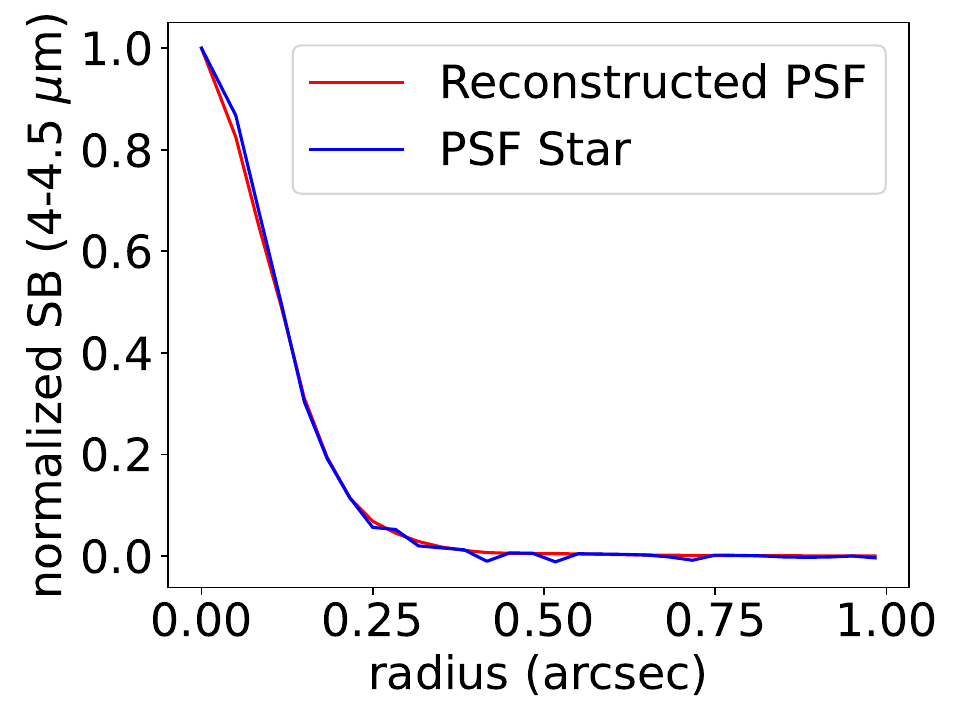}    
\caption{\textbf{Left:} Reconstructed PSF from \qtdfit\ results of J1007$+$2115. The image is integrated over the wavelength range of 4.0-4.5 \mum. \textbf{Right:} Comparison of the azimuthally averaged surface brightness (SB) radial profiles between the reconstructed PSF in the left panel and the empirical PSF constructed from the IFU observation of the PSF star associated with J1007$+$2115.}
\label{fig:fig1}
\end{figure*}

\section{Extended Line Emission from the Quasar Host Galaxy}
\label{sec:4}

\subsection{Quasar PSF Subtraction}
\label{sec:41}

To detect the faint, extended emission from the host galaxy of the quasar, we first carefully remove the bright quasar light from the NIRSpec data cube using \qtdfit\footnote{\url{https://q3dfit.readthedocs.io/en/latest/}} \citep{q3dfit}, a software package designed for the removal of bright point spread function (PSF) from JWST\ data cubes. 
This software is adapted from the well-tested IDL software IFSFIT \citep{rupk17} and has been tested with multiple JWST\ NIRSpec and MIRI IFU observations in previous studies \citep[e.g.,][]{Wylezalek2022, Rupke2023, Vayner2023b, Vayner2023c, VeilleuxLiu2023}. We refer the readers to those papers for more details. Here we summarize the key steps of \qtdfit\ briefly. First, we build a quasar template spectrum from an aperture with a radius of 0.05\arcsec\ (i.e., one spatial pixel or spaxel) centered on the brightest spaxel. Next, for each spaxel within the data cube, we fit the spectrum with a scaled quasar template spectrum representing the quasar PSF contribution in the spaxel ($I^n_{\rm quasar}$), simple featureless monotonic polynomials representing the host continuum emission ($I^n_{\rm starlight, exp.\;model}$), and a set of host emission lines with Gaussian profiles ($I^n_{\rm emission}$). Following the nomenclature of \citet[][]{rupk17}, the decomposition of spectrum in each spaxel described above can be written as $I^n = I^n_{\rm quasar} + I^n_{\rm starlight, exp.\;model} + I^n_{\rm emission}$. 

For the analysis of our object, we have only considered the wavelength range of 3.5 -- 5.0 \mum, where the data quality is good enough and the effect of artificial spectral oscillations is not significant based on visual inspection. Additionally, no host galaxy stellar continuum is detected above 3$\sigma$ level based on our current analysis, so no attempt is made to further characterize the stellar continuum in the rest of this paper.
As for the host emission lines, only \oiiiab\ and \hb\ are considered in the fits as no other emission lines from the host galaxy could be identified within 3.5 -- 5.0 \mum. We have fixed the \oiiiab\ doublet flux ratio to 1:2.98 and {tied the kinematics  (velocities and velocity dispersions) of the corresponding Gaussian components in each emission line together.} We determine our final best-fits by minimizing the reduced chi-square and rejecting any Gaussian components for \oiiitext\ with peak flux density below 2$\sigma$. Our best-fits suggest that one Gaussian component is adequate to describe each of the host emission line profiles. All individual fits are further visually inspected to remove erroneous ones.

The final reconstructed quasar PSF and its comparison with the empirical PSF obtained from the PSF star observation are shown in Fig.\ \ref{fig:fig1}. One caveat for our analysis  is that we assume that the emission from the central $r=0.05$\arcsec\ aperture is purely quasar emission (i.e., no host galaxy contribution). If, instead, the contribution  from the host galaxy is non-negligible in this region, our analysis underestimates the emission from the host galaxy, to a lesser extent on a larger spatial distance away from the quasar.

\begin{figure*}
\includegraphics[width=1.0\textwidth]{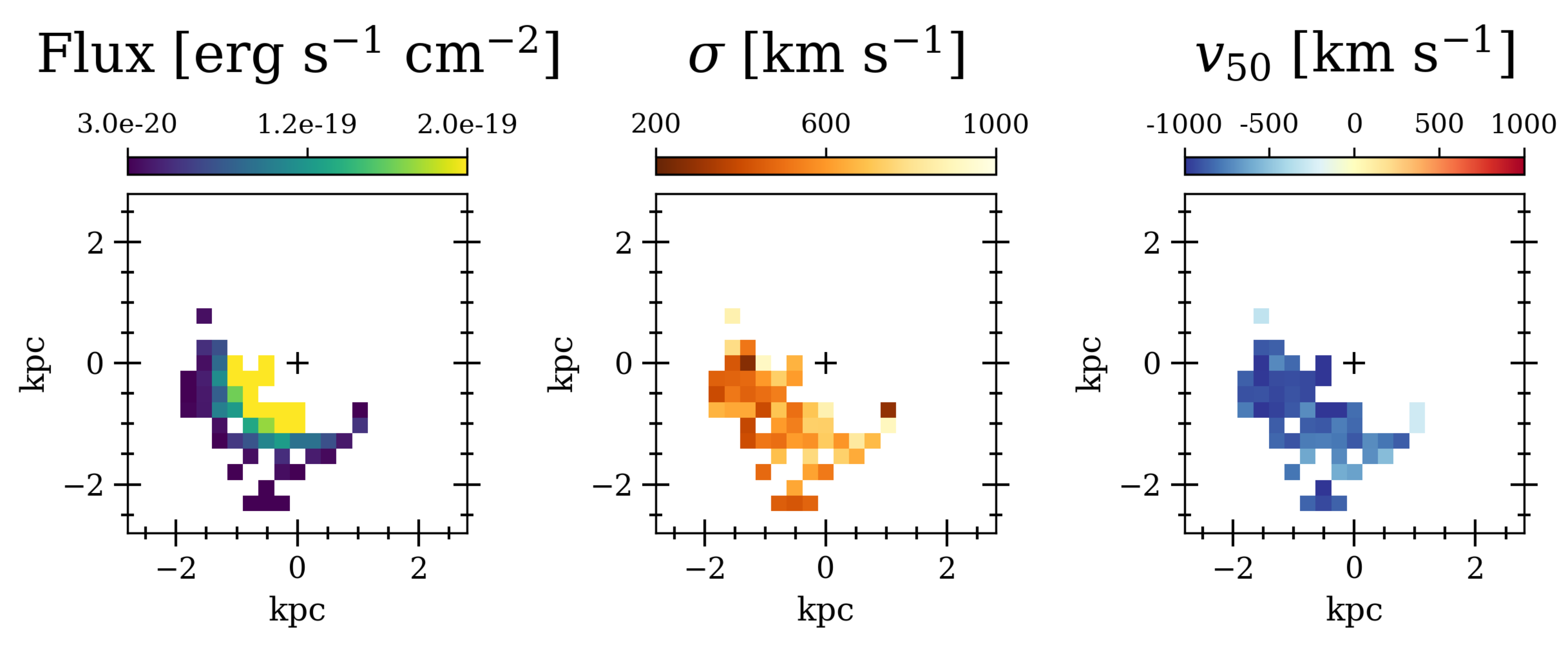}
\caption{Flux, velocity dispersion $\sigma$, and velocity \vwu\ maps of the \oiii-emitting gas in J1007$+$2115, after the quasar PSF has been subtracted. The spatial scales are in unit of kpc. The location of the quasar is indicated by the black cross. {As stated in Sec. \ref{sec:41}, only spaxels with peak flux density above 2$\sigma$ and passing visual inspection are kept and shown in these figures.}  }
\label{fig:fig3}
\end{figure*}

\subsection{Properties of the \oiii\ Emission after PSF Subtraction}
\label{subsec:morphology}

\oiii\ is the strongest emission line in the quasar PSF-subtracted IFU data cube, and a good tracer of the warm ionized gas in quasar host galaxy. The total luminosity of this extended \oiiitext\ emission is $(4.94\pm{0.29})\times10^{42}$ erg s$^{-1}$. In the following sections, we present the observed properties of this \oiiitext\ nebula in detail.

\begin{figure}
\centering
\epsscale{1.3}
\plotone{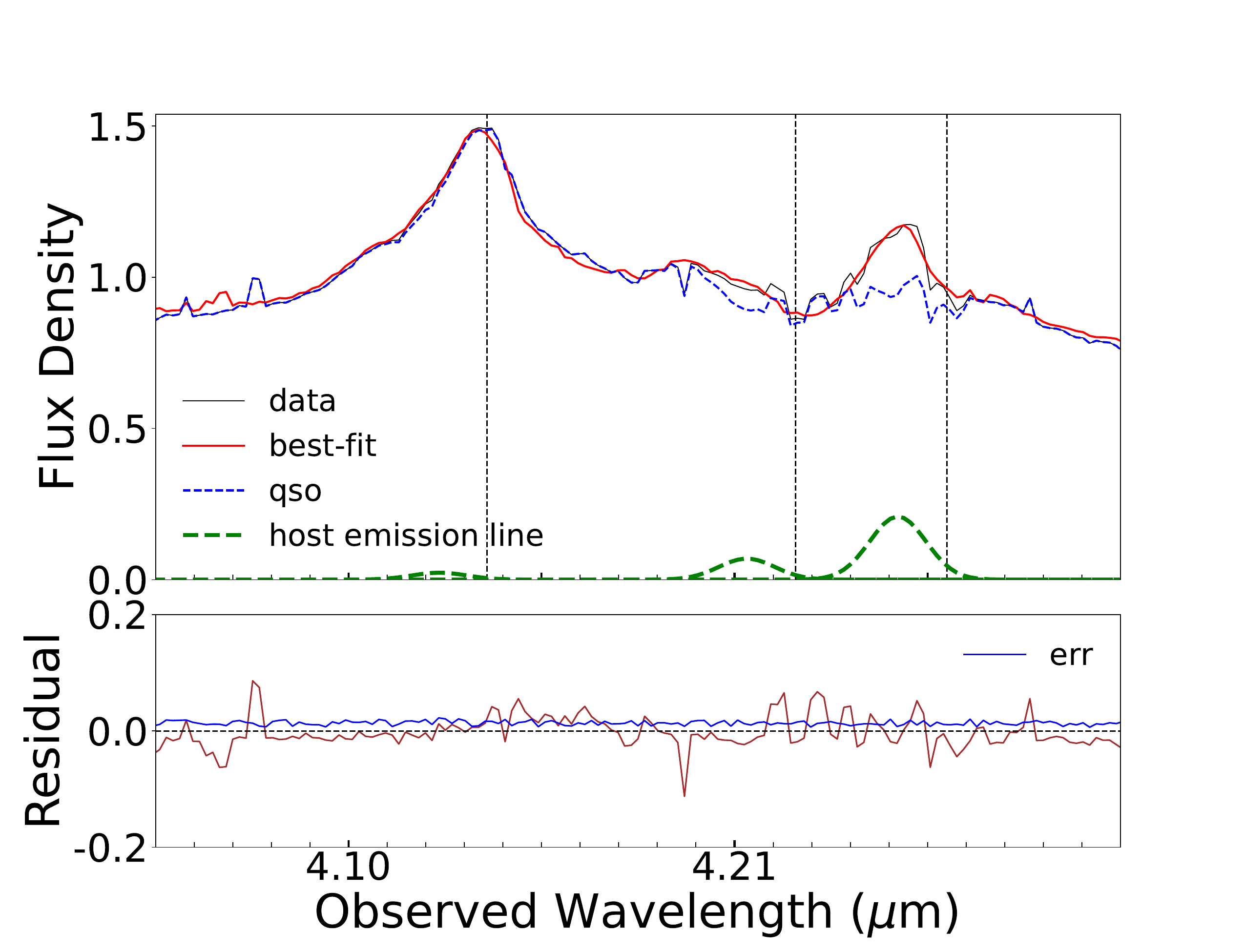}
\caption{An example of best-fit \hb\ and \oiiiab\ emission line profiles from a single spaxel. \textit{Top}: The data in black, the overall best-fit in red, the best-fit scaled quasar template (i.e., the quasar PSF contribution in this spaxel) in blue dashed line and the individual Gaussian components for the host galaxy \hb\ and \oiiitext\ emission lines (from left to right) in green dashed line. The y axis is in arbitrary flux density unit. The location of \hb\ and \oiiitext\ emission lines at systemic velocity are indicated by the vertical dashed lines. \textit{Bottom}: Shown in brown is the difference between the data (black) and the overall best-fit model (red) in the top panel. Shown in blue is the spectral error. The black horizontal dashed line indicates y$=$0.}
\label{fig:example}
\end{figure}

Fig.\ \ref{fig:fig3} shows the emission line flux, velocity dispersion $\sigma$ and radial velocity \vwu\ maps of \oiiitext\ after the quasar PSF has been subtracted by \qtdfit. Here \vwu\ is defined as the velocity at the location where 50\%\ of total line flux is reached. Fig.\ \ref{fig:example} shows an example of \oiiiab\ and \hb\ emission line profiles and their best-fit models from a representative spaxel in the IFU data at $x=-0.56$ kpc and $y =-0.85$ kpc.

The bulk of the \oiiitext\ nebula is located to the southeast of the quasar location, extending over a spatial scale of $\sim$2 kpc to the east and the south with respect to the quasar. The surface brightness of the emission decreases radially. The emission line is highly blueshifted and broad, with radial velocity \vwu\ of $\sim -$260 \kms\ to $-$1150 \kms\ (median: $-$870 \kms) and velocity dispersions $\sigma$ of $\sim$ 260 \kms\ to 920 \kms\ (median: 600 \kms). No signature of a rotating gas component within the host galaxy is found in our data. This line-emitting nebula resembles those tracing quasar-driven outflows at lower redshifts \citep[e.g.,][]{VeilleuxLiu2023}, and we will expand further on this in Section \ref{sec:5}.

\subsection{\oiiitext/\hb\ Line Ratio}

\begin{figure}
    \centering
    \includegraphics[width=\linewidth]{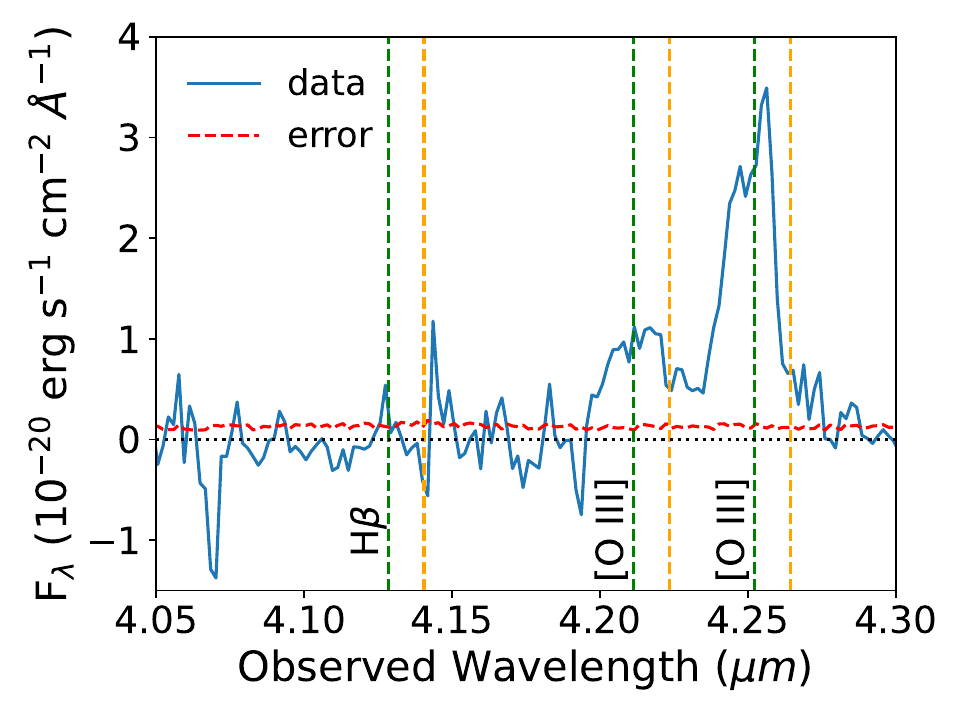}
    \caption{Spectrum stacked over the spaxels wih \oiii\ detections shown in Fig.\ \ref{fig:fig3} from the quasar PSF subtracted data cube. The expected locations of \oiiiab\ and \hb\ emission lines from the outflow at $-$870 \kms\ with respect to the systemic velocity are indicated by the green vertical dashed lines. The orange vertical dashed lines indicate the expected line locations at systemic velocity. The red dashed line represents the error from the original data cube before PSF subtraction.}
    \label{fig:stack}
\end{figure}

After quasar PSF removal, the \hb\ emission is much fainter than the \oiii\ emission, and the measurements from individual spaxels are highly uncertain as their S/N are in general $\lesssim$1. Instead, we stack the spectra from the spaxels where host galaxy \oiii\ emission lines are detected (as shown in Fig.\ \ref{fig:fig3}) from the quasar PSF subtracted data cube, but still obtain an S/N$\lesssim$2 for \hb\ (Fig.\ \ref{fig:stack}). We therefore obtain a 3$\sigma$ upper limit for the \hb\ emission and derive an \oiiitext/\hb\ ratio of $>$10, which is above the typical values seen for low-z star-forming galaxies and is located within the AGN region in the BPT and VO87 diagrams \citep[][]{bpt,Veilleux1987}. It is thus likely that the extended emission detected is photoionized by the quasar, which is usually the case for quasar-driven outflows in Type 1 quasars \citep[e.g.,][]{Hinkle2019}. Nevertheless, some high-z star-forming galaxies also show \oiiitext/\hb\ ratios larger than 10, making the BPT and VO87 diagrams unreliable in the early universe {\citep[e.g.,][and references therein]{Harikane2023,Kocevski2023,Maiolino2023,Sanders2023,Scholtz2023,Ubler2023}}. As a result, we cannot rule out the possibility that the ionization source of the observed extended emission has a stellar origin.

\begin{figure*}
\begin{minipage}[t]{0.33\textwidth}
 \centering
\includegraphics[width=\textwidth]{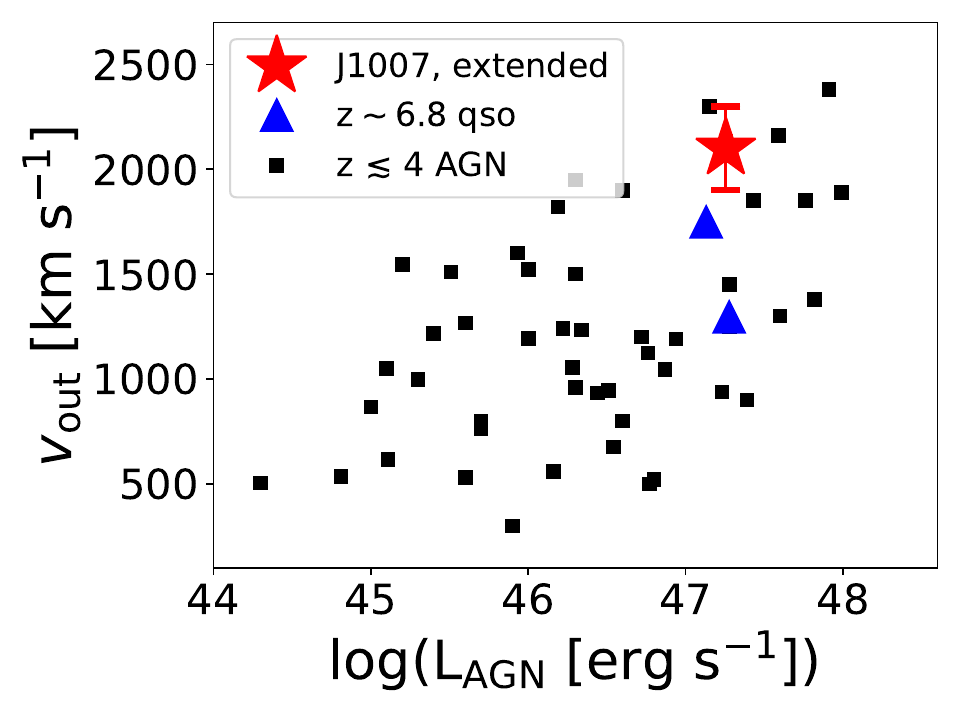}
\end{minipage}
 \begin{minipage}[t]{0.33\textwidth}
 \centering
\includegraphics[width=\textwidth]{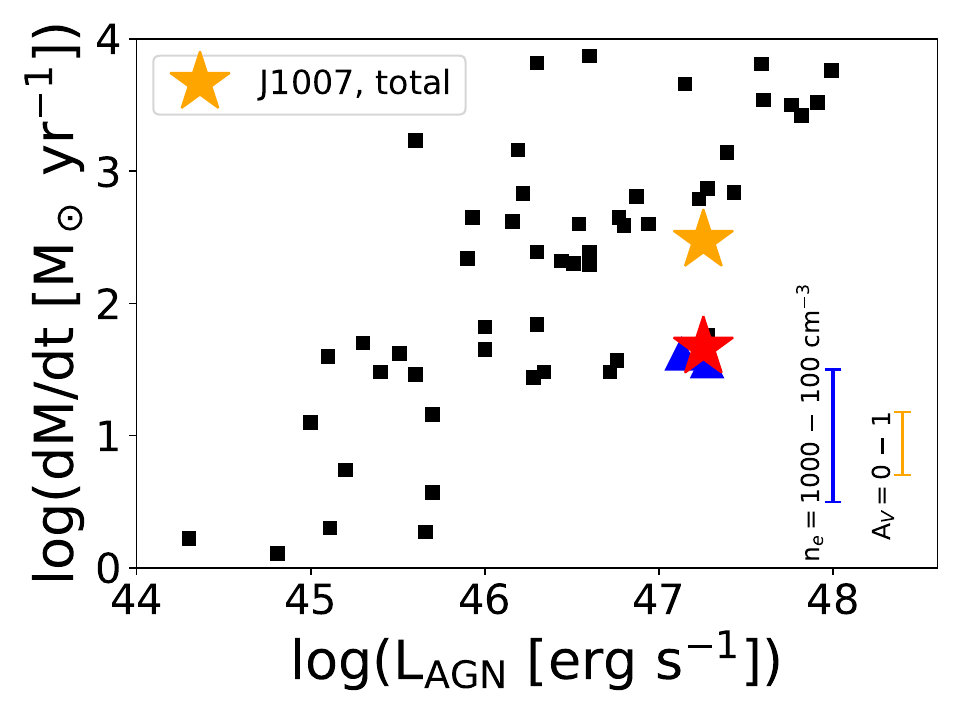}
\end{minipage}
 \begin{minipage}[t]{0.33\textwidth}
 \centering
\includegraphics[width=\textwidth]{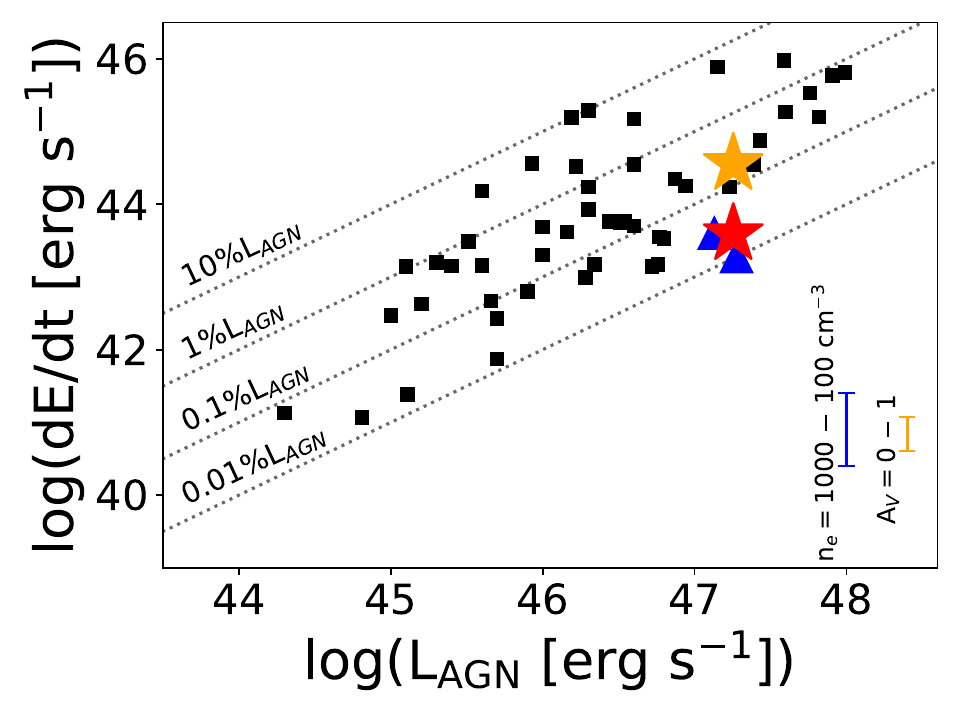}
\end{minipage}
\caption{Maximum outflow velocity (left), mass outflow rate (middle) and kinetic energy outflow rate (right) as a function of bolometric AGN luminosity for the extended outflow in our object J1007$+$2115 (red star) in comparison with the outflows discovered in two $z\sim6.8$ quasars (blue triangles) from \citet{Marshall2023} and other AGN-driven outflows at $z<4$ compiled by \citet{Fiore2017}. In the middle and right panels, for J1007$+$2115, we also show the total rates for both the nuclear and extended outflows combined (orange star). For the $z\sim6.8$ quasars, we only quote the \oiiitext\ emission line based results for consistency. For our object and the $z\sim6.8$ quasars, the error bars are not shown when the sizes of them are smaller than the symbol sizes. For sources from \citet{Fiore2017}, no formal errors were reported for individual objects. {The uncertainties caused by electron density ($n_{e}=\ $1000 -- 100 cm$^{-3}$) and extinction ($A_V =\ $0 -- 1) are indicated by the blue and orange bars at the bottom right corners of the middle and right panels.}}
\label{fig:outflow}
\end{figure*}

\section{Outflow as the Origin of Extended Line Emission}
\label{sec:5}

\subsection{Origin of the Extended Line Emission}

Our \qtdfit\ analysis of the NIRSpec IFU data cube reveals highly blueshifted \oiiitext\ line emission in J1007$+$2115. We discuss the possible origin of this nebula below.  

First, the nebula is unlikely to be gravitationally bounded to the host galaxy. 
The FWHM of the \cii\ emission line from ALMA observations is $\sim$330 \kms\ \citep{Yang2020b} which is in principle an estimate of the characteristic gas velocity within the ISM in this system. It is much smaller than the width of the \oiiitext\ emission (with a median FWHM of $\sim$ 1400 \kms), making the latter highly unlikely to originate from the gas in circular motions. In addition, based on a recent simulation of a luminous quasar at $z\sim7.5$ with similar black hole mass and bolometric luminosity \citep{Ni2018}, the halo mass and circular velocity of our object are on the order of $6\times10^{11}$ \msun\ and 400 \kms, respectively, which is also consistent with other simulations for z$>$6 quasars \citep[e.g.,][]{Costa2015,Costa2018} where the gas velocity rarely reaches above $\sim$500 \kms. The \oiiitext\ has a velocity (median \vwu: $-$870 \kms) larger than this circular velocity and thus, again, does not originate from gas in circular motion. 

Second, there is no evidence that the nebula comes from a merging component or a companion galaxy very close by. The broad linewidth (with a median FWHM of $\sim$1400 \kms) of the emission line is much larger than the typical linewidth seen in quasar companion galaxies at $z>6$. For example, a recent JWST/NIRSpec IFU observation of a quasar-companion merger system reveals much smaller \oiiitext\ linewidths of $\lesssim$250 \kms\ in the entire system (Decarli et al. 2024 in review). Similarly, the typical linewidth of \cii\ emission lines in $z>6$ quasars are also smaller \citep[e.g., an average of $\sim$385 \kms\ for a sample of 27 sources from][]{decarli_alma_2018}. Finally, the typical \oiiitext\ linewidth for star-forming galaxies at $z\sim7$ is also much smaller \citep[e.g., $\lesssim$300 \kms][]{Tang2023}. Therefore, it is unlikely that the broad line emission originates from a merger component or companion galaxy. Note that the large emission linewidth may be explained by a scenario that the system is at a certain phase of merging where a portion of the gas that has been stripped off is falling back, creating large gas velocity dispersion comparable to our observed values. While we cannot completely rule out this scenario, more observations of such quasar merging systems are needed to confirm if they can really reach large linewidth.

The remaining and most likely scenario that explains the large blueshift and broad width of the extended line emission is that it is tracing the fast outflowing gas.
Indeed, the maximum outflow velocity \citep[defined as $v_{out}=v_{50}+2\sigma$ following][]{Fiore2017} is 2100$\pm{200}$ \kms, which falls in the range of quasar-driven outflows at similar quasar bolometric luminosities \citep[e.g.,][left panel in Fig.\ \ref{fig:outflow}]{Shen2016,Perrotta2019,Matthews2021,Yang2023b,Loiacono2024}. Likewise, the broad linewidth also falls in the typical range of outflows in quasars with similar bolometric luminosity \citep[e.g., $\sim$200--2000 \kms;][]{Perrotta2019}. Moreover, the one-sided morphology of this outflow has been seen at lower redshift quasars \citep[e.g.,][]{VeilleuxLiu2023}, where the redshifted (far) side of the outflow might be obscured by the system or intrinsically much fainter. 

Note that, while we cannot formally rule it out, this nebula is improbable to be an inflow located behind the system since (i) it will be much more easily obscured by the galaxy and harder to detect; (ii) the velocity dispersion of an inflow should be much smaller \citep[e.g.,][]{Martin2014}.

\subsection{Energetics of the {\rm \oiiitext} Outflow}
\label{subsec:52}

\subsubsection{Spatially Resolved, Extended Outflow}

The \oiii\ emission line is the brightest line detected from the outflow. 
We thus adopt it to calculate the mass of the outflowing gas \citep{Can2012, Vei2020}:
 \begin{eqnarray}
M_{\rm ionized} & = & 5.3 \times 10^8~\frac{C_e L_{44}([\mathrm{O~III}])}{n_{e,2} 10^{\mathrm{[O/H]}}} M_\odot,
\label{eq:M_ionized}
\end{eqnarray}
where $L_{44}([\mathrm{O~III}])$ is the luminosity of \oiiitext\ normalized to 10$^{44}$ erg s$^{-1}$. In this calculation we assume case B conditions with an electron temperature $T \sim 10^4$ K \citep{Osterbrock2006}. We apply no extinction correction for the \oiiitext\ luminosity due to the lack of constraint on it. The quantity $n_{e,2}$ is the average electron density, normalized to 100 cm$^{-3}$. It is set to 1, a typical value adopted in previous studies of quasar-driven outflows and comparable to the observed values when direct measurements are available \citep[e.g.,][]{Liu2013b,Harrison2014}. Note that the actual dust extinction should be larger than zero (which leads to larger intrinsic \oiiitext\ luminosity and increase gas mass), and the outflowing gas may also have a higher $n_{e,2}$ \citep[which decreases gas mass; e.g.,][]{Harrison16,Jun2020}.  The underestimates of extinction and $n_{e,2}$ have the opposite effects on our estimate. The quantity $C_e \equiv \langle n_e \rangle^2 / \langle n_e^2 \rangle $ is the electron density clumping factor, which can be assumed to be of order unity on a cloud-by-cloud basis (i.e., each gas cloud has uniform density). The quantity 10$^{\rm [O/H]}$ is the oxygen-to-hydrogen abundance ratio relative to the solar value, which is assumed to be 1 (i.e., solar oxygen abundance) in our calculation. 
Based on our \oiii\ measurements, we obtain a gas mass of $(2.6\pm{0.8})\times10^7$ \msun\ for the outflow.

To estimate the total mass, momentum and kinetic energy outflow rates by integrating over the entire outflow, we need to know the dynamical timescale of each parcel $i$ of the outflowing gas, $t_{\rm dyn,i}$ $\approx$ ($R_{\rm deproj,i}/v_{\rm deproj,i}$) = ($R_i/v_i$), where $R_i$ is the measured distance from the center of the gas parcel to the quasar on the sky, and $v_i = |v_{50}|+\sigma$ is the outflow velocity of that same gas parcel. By adopting this formalism for the outflow velocity, we account for the inclination correction needed to recover the true outflow velocity in the 3D space. The linewidth is included as the $\sigma$ term and encodes both the outflow velocity along the line of sight and the turbulent motion of the outflowing gas. Similar approaches have been adopted in previous studies of quasar-driven outflows with NIRSpec IFU \citep[e.g.,][]{Vayner2023c}. The integrated mass, momentum, and kinetic energy outflow rates can thus be written as:

\begin{eqnarray}
\label{eq:Mdot}
\dot{M} & = & \Sigma~\dot{m_i} = \Sigma~m_i~(v_i/R_i) \\
\label{eq:pdot}
\dot{p} & = & \Sigma~\dot{m_i}~v_i \\
\label{eq:Edot}
\dot{E} & = & \frac{1}{2}~\Sigma~\dot{m_i}(v_i)^2 
\end{eqnarray}

\subsubsection{Spatially Unresolved, Nuclear Outflow}

The spectrum extracted from the central spaxel of the data cube resembles the quasar spectrum shown in Fig. \ref{fig:fig0}, where the highly blueshifted \oiiitext\ line emission suggests the existence of a spatially unresolved, nuclear outflow. This is also consistent with the highly blueshifted \civ\ emission reported in \citet{Yang2021}. {However, the central spaxel(s) only represent a portion of the unresolved nuclear outflow as the PSF will distribute the emission to a larger spatial scale beyond the central spaxel(s). The values obtained from the central spaxel(s) would thus be underestimates.} 
Therefore, we then estimate the dynamics and energetics of this spatially-unresolved nuclear outflow. The \oiiitext\ flux from this nuclear outflow is estimated as the difference between the aperture loss-corrected \oiiitext\ flux of the quasar spectrum (i.e., \oiiitext\ flux from the entire outflow) and the \oiiitext\ flux of the extended outflow as shown in Fig. \ref{fig:fig3}. The \oiiitext\ luminosity of the nuclear outflow is then $(1.08\pm{0.08})\times$10$^{43}$ erg s$^{-1}$, the difference between the total \oiiitext\ luminosity from the best-fit in Fig. \ref{fig:fig0}, and the integrated \oiiitext\ luminosity from the extended outflow.
The mass of the nuclear outflow ($M_{nuc}$) is then $(5.7\pm{0.6})\times10^{7}$ \msun\ following Eq.\ \ref{eq:M_ionized} and adopting the same $n_{e,2}$, $C_e$ and 10$^{\rm [O/H]}$ values as in the extended case.
To estimate the mass, momentum and kinetic energy outflows rates, we then follow the same approach adopted for the extended outflow, which are:

\begin{eqnarray}
\dot{M}_{nuc} & = & M_{nuc}~(v_{nuc}/R_{nuc}) \\
\dot{p}_{nuc} & = & \dot{m}_{nuc}~v_{nuc} \\
\dot{E}_{nuc} & = & \frac{1}{2}~\dot{m}_{nuc}(v_{nuc})^2 
\end{eqnarray}

To be consistent with the calculations for the extended outflow, here the outflow velocity is defined as $v_{nuc}=|v|+\sigma$. The size of the outflow $R_{nuc}$ is assumed to be half of the PSF FWHM ($\sim$0\farcs11 or $\sim$0.6 kpc) as measured from the right panel of Fig. \ref{fig:fig1}, which is in principle an order-of-magnitude estimate. As a result, the mass outflow rates derived for the nuclear outflow have large uncertainties.

\subsubsection{Overall Outflow Properties}
The final properties for both the nuclear and extended outflows obtained above are summarized in Table \ref{tab:energetics}. Note that for the mass and outflow rates of the extended outflow, we only list their uncertainties associated with direct measurements (velocity, flux), while the additional uncertainties resulting from the chain of assumptions we made above for other properties (electron density, extinction correction, metallicity, and deprojection) may be an order of magnitude or even larger. {To demonstrate this, the typical uncertainties caused by the two major factors, electron density and extinction correction, are shown in Fig. \ref{fig:outflow}.} In addition, we omit the uncertainties associated with the nuclear outflow rates due to the unknown size of the outflow.

In order to better understand the physical properties of the outflow in our object, we compare its velocity, mass outflow rate, and kinetic energy outflow rate with those of quasar-driven outflows at similar \citep{Marshall2023} and lower redshifts \citep{Fiore2017} with estimates or robust limits on the physical sizes of the outflows based mostly on IFU observations. As shown in Fig.\ \ref{fig:outflow}, the outflow properties of J1007$+$2115 fall within the ranges observed for other quasars with similar luminosities at lower redshifts and follow the general positive trends along with AGN luminosities. If considering only the extended outflow, the mass and kinetic energy outflow rates are on the lower end of the ranges for such quasar outflows. It also suggests that both the launch and impact of the outflow in our object, one of the three earliest quasars at $z>7.5$ known today, is likely similar to those in quasars at later epochs.

In the following sections, we sum up the values for both the nuclear and extended outflows when discussing the dynamics and energetics of the outflow, unless explicitly stated otherwise. We omit the uncertainties for these measurements given the large uncertainties for the nuclear outflow.

\begin{deluxetable*}{c cccc ccc}[!htb]
\tablecaption{Properties of the \oiiitext\ Outflow\label{tab:energetics}}
\tablehead{
& \colhead{$L_{\mathrm{[O III]}}$} &
 \colhead{$V_{\mathrm{max}}$}  &
   \colhead{$R_{\mathrm{out}}$}  & 
  \colhead{$M_{\mathrm{out}}$}  & \colhead{$\dot{M}_{\mathrm{out}}$} & \colhead{$\dot{p}_{\mathrm{out}}$}  & \colhead{$\dot{E}_{\mathrm{out}}$} \\
  & \colhead{[$10^{43}$ erg s$^{-1}$]} &
 \colhead{[\kms]}  &
   \colhead{[kpc]}  & 
  \colhead{[$\times10^{7}$ \msun]}  & \colhead{[\msunyr]} & \colhead{[$\times10^{36}$ dynes]}  & \colhead{[$\times10^{44}$ erg s$^{-1}$]} \\
& \colhead{(1)} & \colhead{(2)} & \colhead{(3)} & \colhead{(4)} & 
\colhead{(5)} & 
\colhead{(6)} &  
\colhead{(7)}
}
\startdata 
{Nuclear} & 1.08$\pm{0.09}$ & $2600\pm{40}$ & 0.6 & $6.8\pm{0.3}$ & $252$ & $3.2$ & $3.2$  \\
Extended & 0.49$\pm{0.03}$ & 2100$\pm{200}$ & 1.5 & $2.6\pm{0.2}$ & 47$\pm{4}$ & $0.48\pm{0.04}$ & $0.39\pm{0.04}$ 
\enddata
\tablecomments{Outflow properties based on the \oiii\ emission line for the spatially unresolved, nuclear outflow (first row) and spatially resolved, extended outflow (bottom row). From left to right, the columns are: (1) \oiiitext\ luminosity; (2) maximum outflow velocity (defined as $|v_{50}|+2\sigma$). For the extended outflow, this is the median value of the entire nebula; (3) median radial distance. For the nuclear outflow, this is the size corresponding to half of the PSF FWHM (i.e., 0\farcs11); (4) mass; (5) mass outflow rate; (6) momentum outflow rate; and (7) kinetic energy outflow rate.  From column (4) to (7), for the unresolved, nuclear outflow, these are order-of-magnitude estimates given the unknown size of the outflow and thus no uncertainties are listed for them. For the extended outflow, we only list the uncertainties associated with direct measurements (flux, velocity), and the additional uncertainties resulting from the chain of assumptions we made for electron density, extinction correction, metallicity and deprojection may be an order of magnitude or even more (see Section \ref{subsec:52} for more details).}
\end{deluxetable*}


\subsection{Power Source of the Outflow}

{The measured momentum outflow rate $\dot{p}$, $\sim 3.7\times 10^{36}$ dynes, is $\sim$61\% the radiation pressure force provided by the AGN}, $L_{\rm AGN}/c \approx (6.0\pm{0.3})\times 10^{36}$ dynes. The quasar is thus capable of driving this outflow via radiation pressure force.

The kinetic energy outflow rate, {$\sim3.6\times10^{44}$} erg\ s$^{-1}$, is $\sim$0.2\% of the bolometric quasar luminosity of our object, $( 1.8\pm{0.1})\times10^{47}$ erg\ s$^{-1}$. This ratio is far below unity but it is still within the range seen in other quasars at lower redshifts \citep[e.g., Fig.\ \ref{fig:outflow}, right panel;][]{rupk17,Har2018}. This again suggests that this quasar can easily drive the observed outflow in a similar manner to other quasars.

We can also examine whether stellar processes are physically capable of driving the observed outflow. At solar metallicity, the typical kinetic energy output rate from core collapse supernovae is $\sim7\times10^{41}(\alpha_{SN}/0.02)(\dot{M_{\star}}/\mathrm{M_{\odot} \ yr^{-1}})$ \citep[e.g.][]{Veilleux2005}. Adopting the SFR of our object (80--520 \msunyr), and assuming a constant supernovae rate of $\alpha_{SN}=0.02$, the expected maximum kinetic energy output rates from core-collapse supernovae in our targets are in the range of $\sim$5.6$\times$10$^{43}$ -- 3.6$\times$10$^{44}$ erg $s^{-1}$. These are {$\sim$ $0.2\times$ -- $1\times$} the kinetic energy outflow rate. Therefore, stellar processes cannot be overlooked as a potential contributing source of energy for this outflow. Note that, however, the kinetic energy output from supernovae as estimated above is based on local relations and may not be applicable at $z\sim7.5$. 

Overall, the observed outflow can be easily driven by the quasar, while the starburst activity may also contribute. 

\subsection{Impact of the Outflow}
\label{subsec:feedback}

The maximum velocity of the extended outflow (2100$\pm{200}$ \kms, the median value of $|$\vwu$|+$2$\sigma$ in the data) is significantly larger than the escape velocity of systems like our object which is on the order of 800 \kms\ according to a recent simulation of quasar-driven outflow at $z\sim7.5$ with comparable black hole mass and bolometric luminosity from the BLUETIDES simulation \citep[][]{Ni2018}. Likewise, the median outflow velocity (870$\pm{70}$ \kms, the median value of $|$\vwu$|$ in the data) is also comparable to the escape velocity.
It is thus likely that the outflow is fast enough to escape the host galaxy and inject energy into the circumgalactic medium and enrich them with metals. This is consistent with predictions from current simulations \citep[e.g.,][and references therein]{Costa2018,Ni2018}, which suggest that kpc-scale outflows with velocities comparable to the one in our object can travel to a scale on the order of 100 kpc. 

Additionally, this outflow may also help clear out gas and dust along the way and thus make it easier for the quasar radiation to escape the galaxy \citep[e.g.,][]{Costa2018,Bennett2024}, and help with the ionization of the gas around the quasar. Likewise, this also helps with the formation of \lya\ nebulae usually associated with such quasars \citep[e.g.,][]{Costa2022}.

Combining both the nuclear outflow and extended outflow together, the mass outflow rate, {$\sim$300 \msunyr, in J1007$+$2115 is $\sim$60\%--380\% of the SFR (80--520 \msunyr)} inferred from the \cii\ luminosities from ALMA observations \citep{Yang2020b}. While the gas consumption in this system is still likely dominated by the star formation, the outflow is able to expel a significant amount of gas from the inner part of the host galaxy. 
The ratio of kinetic energy outflow rate to quasar bolometric luminosity is {$\sim$0.2\%, which is comparable to the minimum value ($\sim$0.1\%) necessary for outflows to provide negative feedback to their host galaxies as predicted by some simulations \citep[e.g.,][]{Choi2012,Hopkins2012}.} However, it is worth noting that there are other simulations requiring higher ratios ($>$0.5\%) to allow for negative feedback \citep[see][for a recent compilation of such simulation predictions]{Harrison2018}.

Furthermore, in such luminous quasars, the outflows are expected to be multi-phase, and our results have not accounted for potential highly-ionized, neutral, or molecular outflows in this object, with the latter two phases usually containing much more mass and energy \cite[e.g.,][]{Veilleux2020}. The observed mass and kinetic energy outflow rates obtained above may be lower limits to the total outflow rates in our object, and the feedback provided by the outflow on the host galaxy may thus be underestimated.

The outflow detected in our object, one of the three earliest quasars known, represents the onset of quasar feedback that may be responsible for the quenching of passive/quiescent galaxies at $z\sim$\ 2-5 \citep[e.g.,][]{Labbe2005, Glazebrook2017,
Valentino2020,Valentino2023,Alberts2023,Ji2024,Nanayakkara2024}. For example, GS-9209, a massive quiescent galaxy at $z=4.658$, could be quenched as early as $z\gtrsim6.5$ \citep{Carnall2023}, the same era that our object lives in.
The outflow detected in J1007$+$2115 may just represent the very early phase of quasar feedback, and multiple outflow events may occur as the system evolves and quench/regulate the star formation activity within the galaxy.

\subsection{Constraints on the Quasar Lifetime from the Outflow Time-scale.}
We can estimate the dynamical timescale for the outflow to travel from the quasar to its current location. Adopting the median outflow radial distance of $\sim$1.5 kpc (the median value of the radial distance of individual spaxels in Fig.\ \ref{fig:fig3}) and the median outflow velocity of $\sim$870 \kms, we obtain a timescale of $\sim$1.7 Myr. Therefore, as the outflow is launched by the quasar, the quasar itself should be active at least $\sim$1.7 Myr ago. 

In order for the $\sim$10$^{9}$ \msun\ black hole in our object to grow this massive at $z\sim7.5$, it requires continuous accretion at the Eddington limit since $z\sim30$ assuming a radiative efficiency of 0.1 and a seed black hole mass of 10$^4$ \msun\ \citep[e.g.,][]{Yang2020b}. The fraction of time that the quasar has been inactive since $z\sim30$ is likely very small. As a result, we can safely assume that the quasar continues to be active since the onset of the outflow, and estimate the quasar lifetime \TQ\ to be $\gtrsim$1.7 Myr in this scenario. 
This lower limit on the quasar lifetime falls within the range of quasar lifetime for early quasars at $z\gtrsim6$ (e.g.,$\lesssim10^8$ yr) based on proximity zone studies \citep[e.g.,][]{Eilers2017} and IGM damping wing studies \citep[e.g.,][]{Davies2018b,Wang2020,Durovckova2024}.

\section{Summary}
\label{sec:6}

In this paper, we examine in detail the deep JWST\ NIRSpec/IFU data of quasar J1007$+$2115  at $z=7.5149$, one of the earliest luminous quasar yet known. 
Our main results are summarized below.

\begin{itemize}

\item
We obtain a bolometric luminosity of $(1.8\pm{0.1})\times10^{47}$ \ergs\ based on the 5100 \AA\ continuum luminosity. {For black hole mass, we obtain $(7.0\pm{0.4}) \times10^{8}$ \msun\ or $(2.5\pm{0.2}) \times10^{9}$ \msun\ based on the \hb\ adopting the quasar spectrum, if we include or exclude the narrowest Gaussian component of best-fit \hb\ emission line profile. These results lead to an Eddington ratio of $2.2\pm{0.1}$ or $0.6\pm{0.1}$).} A faint but highly blueshifted \oiii\ emission line is also present in the quasar spectrum. 

\item
An extended \oiiitext\ line-emitting nebula is detected by the JWST\ NIRSpec/IFU data after the quasar PSF has been subtracted. This line emission is highly blueshifted and broad, and extends to $\sim$2 kpc away from the quasar. The emission is most likely tracing a rapid outflow in this quasar host galaxy, and is the earliest galactic-scale outflow known at present.

\item
We obtain \oiiitext/\hb\ $>$ 10 based on the stacked spectrum of the extended nebula from the PSF-subtracted data cube, adopting the 3$\sigma$ upper limit on the flux of \hb\ emission. This suggests that the outflowing gas is dominated by AGN photoionization.

\item 
In addition to the extended \oiiitext\ outflow, there is also a spatially unresolved, nuclear \oiiitext\ outflow in our object. For both the nuclear and extended outflows, the velocity, mass outflow rate, and kinetic energy outflow rate fall within the ranges observed for other outflows in quasars at lower redshifts, although the latter two are at the lower end of the measured ranges when only considering the extended outflow.

\item 
Combining the nuclear and extended outflows, the momentum outflow rate, {$\sim3.7\times10^{36}$ dynes, is $\sim$61\%} of the radiation pressure force provided by
the quasar, and the kinetic energy outflow rate, {$\sim3.6\times10^{44}$} erg s$^{-1}$ is {$\sim$0.2\%} of the quasar bolometric luminosity. These suggest that the quasar is powerful enough to easily drive the outflow. However, we cannot formally rule out the possibility that the starburst activity also contributes to the launch of the outflow.

\item 
The large velocity of the extended outflow ($\sim$2100 \kms) implies that it may easily escape the host galaxy (with escape velocity on the order of 800 \kms). The outflow may thus help inject energy to the circumgalactic medium and enrich it with metals. Similarly, the outflow may also clear out the dust and gas along the way, which helps with the escape of quasar radiation.

\item 
Combining the nuclear and extended outflows, the mass outflow rate, {$\sim$300 \msunyr, is $\sim$60\%--380\% of the SFR (80--520 \msunyr)}. While the star formation activity may still dominate the gas consumption, the outflow is capable of expelling a significant amount of gas from the inner region of the host galaxy. The ratio of kinetic energy outflow rate to quasar bolometric luminosity, $\sim$0.2\%, is comparable to the minimum value ($\sim$0.1\%) required for negative quasar feedback on the host galaxy according to some simulations.

\item
The average dynamical timescale of the extended outflow is estimated to be $\sim$1.7 Myr. This sets a lower limit for the lifetime of this quasar, which is consistent with the quasar lifetime obtained from proximity zone studies and damping wing studies of similar quasars at $z>6$.

\end{itemize}

\begin{acknowledgments}
We thank the anonymous referee for constructive comments that have improved the paper. W. L. thanks James Davies for his kind help with the software snowblind.
W. L., X. Fan, J. Yang, were supported in part by NASA through STScI grant JWST-GO-1764. E.P.F. is supported by the international Gemini Observatory, a program of NSF NOIRLab, which is managed by the Association of Universities for Research in Astronomy (AURA) under a cooperative agreement with the U.S. National Science Foundation, on behalf of the Gemini partnership of Argentina, Brazil, Canada, Chile, the Republic of Korea, and the United States of America. A. L. acknowledges support by the PRIN MUR “2022935STW". F. L. acknowledges support from the INAF GO 2022 grant "The birth of the giants: JWST sheds light on the build-up of quasars at cosmic dawn" and from the INAF 2023 mini-grant "Exploiting the powerful capabilities of JWST/NIRSpec to unveil the distant Universe". This work is based on observations made with the NASA/ESA/CSA James Webb Space Telescope. The data were obtained from the Mikulski Archive for Space Telescopes at the Space Telescope Science Institute, which is operated by the Association of Universities for Research in Astronomy, Inc., under NASA contract NAS 5-03127 for JWST. These observations are associated with program \#1764. 

\end{acknowledgments}

%

\vspace{5mm}
\facilities{JWST}

\software{astropy \citep{Ast2013, Ast2018}, reproject (\url{https://doi.org/10.5281/zenodo.7584411}), \qtdfit\ \citep{q3dfit}}





\bibliography{J1007}{}
\bibliographystyle{aasjournal}



\end{document}